\documentclass{iaus}
\usepackage{graphicx}
\usepackage[authoryear]{natbib}

\title[ZDI : old problems and new methods] 
{Zeeman-Doppler Imaging : old problems and new methods}

\author[Carroll et al.]   
{T.A. Carroll$^1$, M. Kopf$^1$, K.G. Strassmeier$^1$, I. Ilyin$^1$}

\affiliation{$^1$Astrophysikalisches Institut Potsdam,
An der Sternwarte 16, D-14882 Potsdam, Germany \break email: tcarroll@aip.de}

\pubyear{2009}
\volume{259}  
\pagerange{119--126}
\date{?? and in revised form ??}
\jname{Proceedings Title IAU Symposium}
\editors{A.C. Editor, B.D. Editor \& C.E. Editor, eds.}
\begin{document}

\maketitle

\begin{abstract}
Zeeman-Doppler Imaging (ZDI) is a powerful inversion method to reconstruct
stellar magnetic surface fields.
The reconstruction process is usually solved by translating the
inverse problem into a regularized least-square or optimization problem. 
In this contribution we will emphasize that ZDI is an inherent non-linear problem 
and the corresponding regularized optimization is, like many non-linear problems, 
potentially prone to local minima. We show how this problem will be exacerbated 
by using an inadequate forward model. 
To facilitate a more consistent full radiative transfer driven approach to ZDI we 
describe a two-stage strategy that consist of a
principal component analysis (PCA) based line profile reconstruction 
and a fast approximate polarized radiative transfer method to synthesize 
local Stokes profiles. 
Moreover, we introduce a novel statistical inversion method 
based on artificial neural networks (ANN) which provide a fast
calculation of a first guess model and allows to incorporate
better physical constraints into the inversion process. 

\end{abstract}

\firstsection 
\section{Introduction}

Zeeman-Doppler Imaging (ZDI) was pioneered by \cite{Semel89}, although
first described as a mere detection method for surface magnetic fields on rapid rotating stars, 
the method quickly evolved to a real mapping (inversion) technique \citep{Donati90,Brown91}.
Today ZDI or Magnetic-Doppler Imaging (MDI) is a synonym for the inversion of time and phase resolved 
spectropolarimetric observations to reconstruct stellar magnetic
surface fields. Since that time ZDI was a success story of its own and has enormously 
contributed to our current understanding of stellar magnetism \citep[e.g.][]{Donati97,Donati99,Hussain02,
Oleg04,Donati06,Donati07,Petit08}.
However, there are a number of fundamental problems which has to be considered
when dealing with ZDI as an inverse problem. One of these problems comes from
an inadequate formulation and approximation of the forward problem while another comes from the 
inherent non-linearity of the inverse problem.

\section{The forward problem}
\label{Sect:2}

Inverse problems arise whenever one searches for causes of observed data or desired
effects. The most general way of solving the inverse problem (directly or indirectly) 
is to utilize a model of the underlying forward problem. 
The forward problem constitutes our physical theory and/or knowledge about the system 
to make predictions about the outcome and result of measurements. 
The direct or forward problem in ZDI is typically the polarized radiative transfer based 
on a model of the stellar atmosphere.
Here we may encounter the first fundamental problem, owing to our limited 
theoretical knowledge (e.g. lack of atomic and/or atmospheric parameters) 
as well as numerical and computational constraints we are restricted 
to an approximate description of the \emph{real} physical system. 
Critical in this context is to provide an appropriate and adequate 
parameterization of the model , i.e the minimal number of free variable
that determine the measurable outcome. The determination of a particular 
parameterization is not always obvious and requires a careful evaluation.   
Formally the forward problem of ZDI can be written in a compact way, by using the 
formal solution of the Stokes vector as described by \cite{Landi85}. This allows us 
to write the disk-integrated Stokes vector $\boldsymbol{I}^{*}$
at a wavelength $\lambda$ and for a particular rotational phase $\phi$ as follows :
\begin{equation}
\label{Eq:1}
 \boldsymbol{I}^{*}(\lambda,\phi)  = \int_{\hat{M}} \int_{0}^{\infty}
 \boldsymbol{O}\left ( 0,s(M',\theta),\boldsymbol{X}(M'),M',\phi \right )
 \; \boldsymbol{j} \left ( s(M',\theta),\boldsymbol{X}(M'),M',\phi \right ) ds \cos\theta  dM .
\end{equation}
The vector $\boldsymbol{X}$ provides the parameterization of the model and 
comprises parameters like the magnetic field vector, temperature, abundance, etc., 
$dM$ denotes the infinitesimal surface element at the position $M'$ (given in spherical 
coordinates), $\theta$ denotes the angle between the surface normal at $M'$ and the direction 
to the observer, $s$ is the geometrical path length towards the observer, 
$\boldsymbol{j}$ is the emission vector \citep[see][]{Stenflo94}, and $\boldsymbol{O}$ is the 
evolution operator of the formal solution which incorporates the absorption matrix. Following 
\cite{Landi85}, we can write the evolution operator, in the case of a piecewise constant 
absorption matrix $\boldsymbol{K}$, as
\begin{equation}
\label{Eq:2}
\boldsymbol{O}(s,s',M,\boldsymbol{X}(M)) = e^{- \boldsymbol{K}(M,\boldsymbol{X}(M)) \; |s'-s|} \; .
\end{equation} 
Equation \ref{Eq:1} can also be written symbolically in a more compact form as
\begin{equation}
\label{Eq:3}
\boldsymbol{I}^{*}(\lambda,\phi) \; = \; \boldsymbol{\Omega}(\boldsymbol{X}) \; ,
\end{equation}
where $\boldsymbol{\Omega}$ represents the formal non-linear integral operator of the forward
problem. Several things must be kept in mind here, 
both integrals, the outer surface integration over the model atmosphere as well as the inner 
transport integral through the atmosphere, 
must be discretized in order to cope with the problem numerically. We should be aware that this 
already means that we transform our originally continuous problem into a discrete one 
with finite dimension. As mentioned above, the finite parameterization of our model $\boldsymbol{X}(M)$ 
deserves particular attention because the limitation to those 
variables that we consider as necessary and sufficient, will directly determine the dimension
of the model space, we will see later how the neglect of the temperature may have drastic
consequences for the inversion.
The radiative transfer model we use is still formulated in local thermodynamic equilibrium 
(LTE) which is quite sufficient for the majority of photospheric spectral lines we are using but 
certainly has its limitations. Moreover, it is customary to describe the atmosphere as
height independent, i.e. neglecting all height gradients in most of the model parameter, 
an approximation which becomes questionable in particular for stars 
with extended atmospheres like giants.
All these approximations of the \emph{real} physical situation 
in stellar atmospheres will therefore have direct consequences on the inversion process.
Finally, it should be also clear from equation (\ref{Eq:1}) and (\ref{Eq:2}) that our forward 
problem is in general non-linear in $\boldsymbol{X}$, an aspect that will be discussed 
further in the following section.

\section{The inverse problem}
\label{Sect:3}
The retrieval of model parameters from measured data 
constitutes the general definition of a inverse problem.
In the particular case of ZDI the measured data are given in the form of phase resolved
Stokes profile observations. In general there is a
distinction between continuous and discrete inverse problems which means that we 
want to retrieve either a continuous functions or a parameterized model. 
In that sense it is already the formulation of the forward problem which defines the kind of inverse problem.
The same holds for the definition of linear or non-linear inverse problems, which is determined 
by the linearity or non-linearity of the forward problem. 
Inverse problems, like ZDI, are often said to be ill-posed by which, in general, we mean that
i.) the inverse problem might be non-unique and ii.) the inverse problem might not depend 
continuously on the data, i.e. the solution is unstable against small perturbation of the observed data.
The non-uniqueness may have different reasons which can be a real intrinsic non-uniqueness of the problem, 
or uncertainties in the observed data, or an ill-defined forward model, or a 
lack of data that sufficiently constrain the problem. 
The problem of ill-posedness can be effectively addressed by a
regularization approach, i.e. adding additional a priori assumptions (i.e. constraints) to
the inverse operator. A powerful method in this context is the Tikhonov
regularization but many other forms of regularizations are also possible.
For linear problems it can indeed be shown that the regularization
leads to a unique and stable solution \citep[see e.g.][]{Engl96}.
However, the forward problem we deal with in ZDI is non-linear, and at first, it is not obvious
how this will affect the uniqueness and stability of our solution.
The general way to proceed
is to formulate and minimize a regularized least-square problem.
This approach, relies on the linearization of the forward operator,
which can be obtained in our case from a first order expansion of (\ref{Eq:3}),
\begin{equation}
\label{Eq:4}
\boldsymbol{\Omega}(\boldsymbol{X}) \; = \; \boldsymbol{\Omega}(\boldsymbol{X}_0) + 
\boldsymbol{J}(\boldsymbol{X}-\boldsymbol{X_0}) \; ,
\end{equation} 
where $\boldsymbol{J}$ is the Jacobian matrix of $\boldsymbol{\Omega}$. Using the sum of square error and a regularization
functional $G(\boldsymbol{X})$ the objective functional $F(\boldsymbol{X})$ can be written as
\begin{equation}
\label{Eq:5}
F(\boldsymbol{X}) \; = \; \|\boldsymbol{I}_{obs} - \boldsymbol{I}^{*} \|^2 \; + \; \alpha G(\boldsymbol{X})
\end{equation}
where $\alpha$ is the regularization parameter which determines the influence of the regularization. 
The regularization functional $G$ can be expressed in the from
$\|L(\boldsymbol{X}_{ref}-\boldsymbol{X}_0) \|^2$ where $\boldsymbol{X}_{ref}$ is a parameter 
vector which represents our a priori knowledge (assumption) about the solution, and $L$ is a suitable matrix
approximation of a differential operator. 
The essence of the non-linear solution is then
to repeat the process of solving the linearized solution in terms of a gradient descent method. 
Using a Gauss-Newton type form of linearization \citep{Engl96}
we obtain from (\ref{Eq:5}) for an increment $\Delta \boldsymbol{X}_{n+1}$ at the iteration $n+1$ the following
solution, 
\begin{equation}
\label{Eq:6}
\Delta \boldsymbol{X}_{n+1} \; = \; \left ( \boldsymbol{J}^* \boldsymbol{J} + \alpha \boldsymbol{L}^*\boldsymbol{L} \right )^{-1} 
\left (\boldsymbol{J}^*(\boldsymbol{I}_{obs} - \boldsymbol{I}^{*}) \right ) + 
\alpha \boldsymbol{L}^*\boldsymbol{L}(\boldsymbol{X}_{ref}-\boldsymbol{X}_n) \; ,
\end{equation} 
where $\boldsymbol{J}^*$ is the conjugate transpose of $\boldsymbol{J}$.
The stabilizing effect of the regularized solution is then ensured by the positive
definiteness of the inverse
$( \boldsymbol{J}^* \boldsymbol{J} + \alpha \boldsymbol{L}^*\boldsymbol{L} )^{-1}$. 
As the regularization parameter is held constant throughout the iteration, the regularization part in (\ref{Eq:5}) 
will have an increasing influence as we approach the solution and thus provides the additional benefit of preventing 
the inversion from fitting arbitrary  details (i.e. noise) in the spectra.
However, although we gain a stabilizing effect at each iteration step,
it should be clear from (\ref{Eq:5}) that even when $G$ provides a convex constraint this needs 
not necessarily be true for $F(\boldsymbol{X})$ which, in general, is still a non-convex function 
due to the non-linearity of (\ref{Eq:1}). 
The inverse problem may therefore have many
local minima in the parameter space in which a gradient descent method can get stuck and 
a regularization alone can not guarantee to find the unique global minimum. 
Furthermore, the determination of an 
appropriate regularization parameter $\alpha$ is by no means an easy task and can require
high computational efforts \citep{Engl96}.

\section{More than one solution -- an inadequate forward model}
\label{Sect:4}
\begin{figure}
 \includegraphics[width=5.7cm,height=11cm]{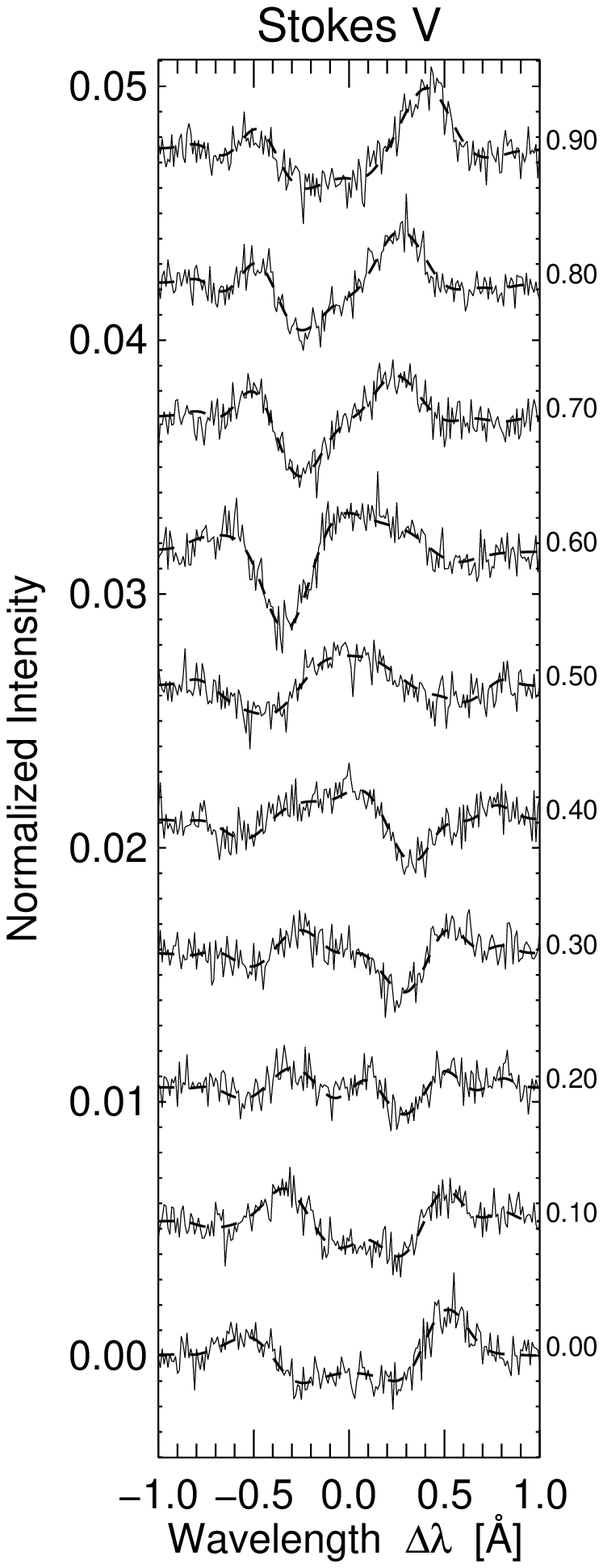}
 \includegraphics[width=5.7cm,height=11cm]{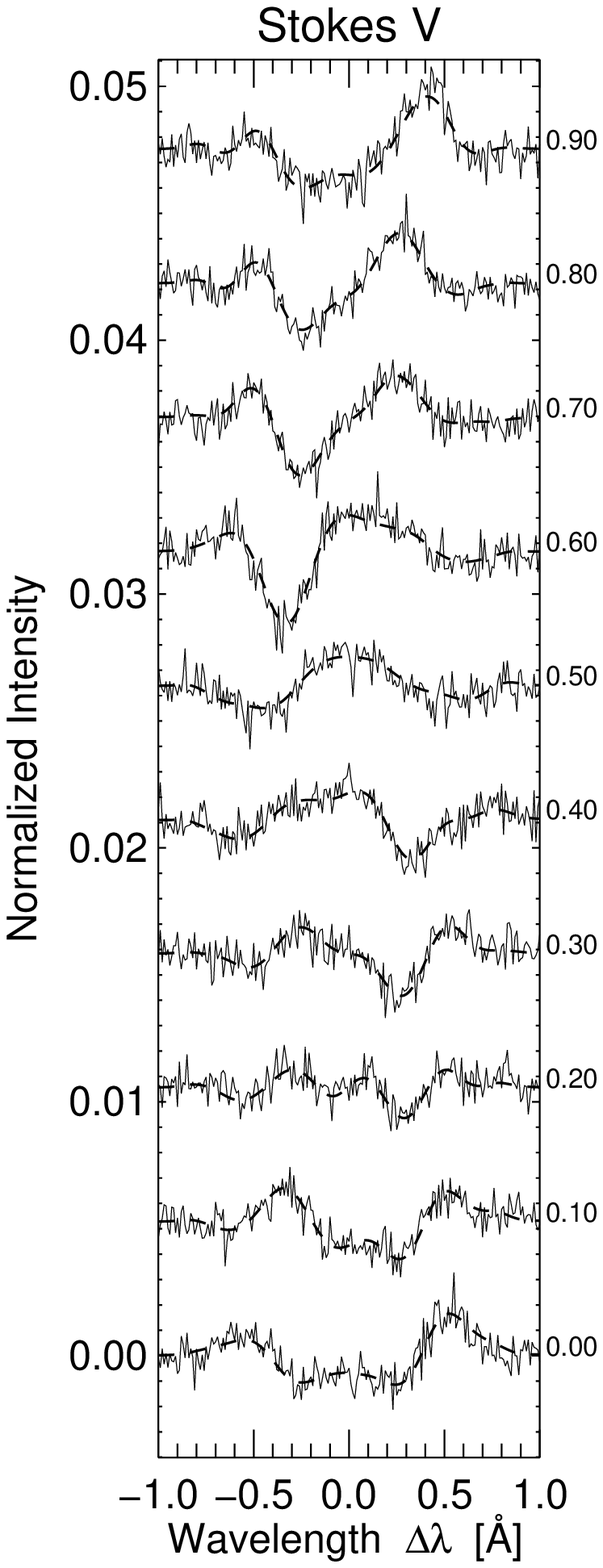}
  \caption{The fits (dashed lines) to the noisy reference Stokes $V$ profiles from the two test inversions : 
  on the left the fits of the simultaneous temperature and
  magnetic field inversion, $\chi^2_{red}$ = 1.01 and on the right the fits of the restricted 
  inversion, $\chi^2_{red}$ = 1.02.}  
  \label{Fig:2}
\end{figure}

To show how much the solution of the ZDI problem depends on the forward model and the chosen 
parameterization we made a simple synthetic experiment. The stellar model surface 
consists of a large bipolar magnetic field region. 
The stellar atmosphere is that of a typical fast rotating K subgiant star (i.e. log$(g) = 3.5$,
$T_{eff}$ = 4600 K, [M/H] = -0.5, vsini = 40 km/s).
One of the magnetic regions is located in a cool spot like structure while the other polarity
is embedded in a hot and more dispersed like structure to mimic a faculae or plage region, see top 
of Figure \ref{Fig:2}
The magnetic field is exclusively radial oriented in both magnetic regions. 
Two inversion were performed, the first one worked on the basis of the correct parameterization with the magnetic field vector and
temperature as free parameter, while the second inversion runs with a fixed temperature. This allows the
second inversion only to seek in a subspace for a possible solution. For the inversion we have our
ZDI code \emph{iMap} \citep{Carroll07} with a surface segmentation of 1800 elements, and the
Zeeman-sensitive iron line FeI $\lambda$ 6173.
Although we have all four Stokes components available we restricted the number of available Stokes components 
to Stokes $I$ and Stokes $V$ of the Zeeman-sensitive iron line FeI $\lambda$ 6173. We have added noise with a signal-to-noise
level of 1000 to the original model spectra. Both inversion run with the same value for the regularization parameter.
\begin{figure}
 \includegraphics[width=13cm]{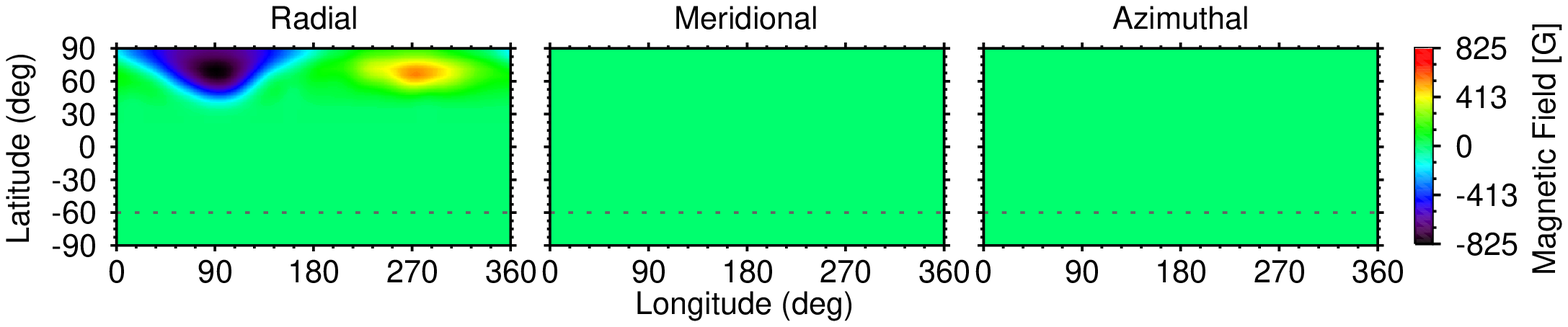}
 \includegraphics[width=13cm]{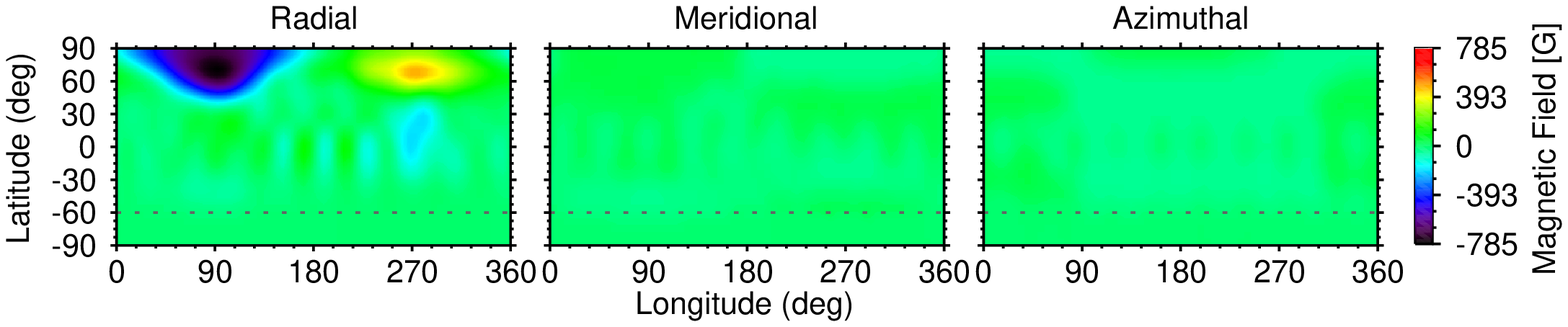}
 \includegraphics[width=13cm]{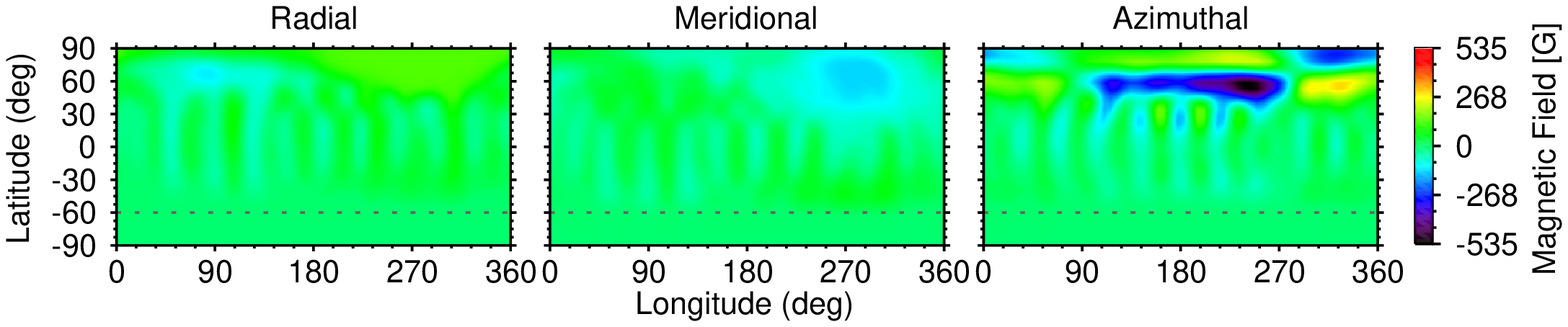}  
  \caption{The mercator plot of the magnetic field distribution : (top) the synthetic reference model, (middle) the
           reconstructed magnetic field from a simultaneous temperature and magnetic field inversion, (bottom) the
	   result from the restricted inversion (fixed temperature).}  
  \label{Fig:1}
\end{figure}
The first inversion converged to a solution that is close to the reference field structure, see Figure \ref{Fig:1}
middle. The surface structure is well reconstructed from the simultaneous temperature and magnetic field inversion 
and also the Stokes $V$ spectra (Figure \ref{Fig:2}, left) are well reproduced.
In the second inversion where we have neglected the temperature as a free parameter the fit to the reference spectra
is again very good but the resulting magnetic surface field is drastically different (Figure \ref{Fig:1}, bottom).
The surface field shows a characteristic high latitude band of a 
unipolar azimuthal field with an longitudinal extension of about 180$^0$ which is in sharp contrast 
to the original reference field. The inversion in the restricted parameter regime has apparently 
converged to a false solution, although the fits are very good (Figure \ref{Fig:2}, right). 
The quality (goodness) of the fit (Figure \ref{Fig:2}) itself do not 
provide enough information in this case to distinguish between the competing solutions. 
This issue could readily be resolved if the corresponding Stokes $Q$ and $U$ profiles are 
taken into account (not shown here).
Due to the critical role of the temperature and the associated radiative transfer effects
it will be of particular importance to provide an accurate and appropriate forward model for the inverse problem. 
We will investigate the impact of radiative transfer effects and inadequate modeling for ZDI 
inversions in greater detail in a forthcoming paper \citep{Carroll09}.

\section{Towards a full polarized radiative transfer approach}
\label{Sect:5}
We have seen from the foregoing synthetic example how the underlying model can affect the 
outcome of the inversion. Moreover we have seen the critical role of the temperature which is
of course tightly correlated to the radiative transfer effects. It is therefore worthwhile to 
include a full polarized radiative transfer in the inversion process to make the the problem setting
as realistic as possible. There are at least two major obstacles towards a full radiative transfer driven 
approach :
\begin{itemize}
\item the noise level of individual Stokes line profiles, and
\item the computational demands of an iterative inversion.
\end{itemize}
The first point refers to the cases that most observed and measured Stokes $V$, $Q$ and $U$ spectra are heavily
contaminated by noise. Although there exist powerful multi-line reconstruction techniques as the least-square
deconvolution method of \cite{Donati97}, they do not easily allow one to apply a radiative transfer modeling approach.
The crucial point is that this technique only
retrieve a kind of \emph{mean} line profile which has no straight forward interpretation
on the basis of spectral line synthesis.
The second point highlights the computational effort one is facing by the iterative approach to
the inverse problem. 
To address both of the above listed problems and to pave the way for a full
radiative transfer modeling approach we propose a two-step strategy, a 
multi-line reconstruction method which enables the reconstruction of individual Stokes profiles 
and a fast approximate radiative transfer calculation by artificial neural networks.

\subsection{Multi-Line PCA reconstruction}
\label{Sect:5.1}
The basic idea of the multi-line PCA reconstruction technique is to  
extract the systematic features within the observed data set, i.e. Stokes profiles,
and at the same time to reduce the noise.
This is done by identifying the redundancy within the dataset and describing the dataset with a
minimal number of characteristic parameter (i.e. dimensions) as possible. 
The redundancy among the observed data can be described 
by the correlation between the individual spectra such that we can express the correlation 
of entire observational dataset by the covariance matrix $\boldsymbol{C}_x$ which can be written, 
after transforming the spectra into the velocity domain, as
\begin{equation}
\boldsymbol{C}_x = \sum_n \left ( \boldsymbol{X}_n(v) - \bar{\boldsymbol{X}}(v) \right ) \left ( \boldsymbol{X}_n(v) - \bar{\boldsymbol{X}}(v) \right )^T \; ,
\label{Eq:7}
\end{equation}
where $v = c \Delta\lambda/\lambda$ and $n$ is the number of individual spectral line profiles used for
the analysis and $\bar{\boldsymbol{X}}$ the mean Stokes profiles of all spectral lines.
We then seek a set of orthonormal direction in the data space (given by the spectral resolution) along which 
the variance in $\boldsymbol{X}$ is maximized in descending order.
These new set of coordinate axes which accounts for the maximum variance in the observed data (Stokes spectra)
can be determined by calculating the eigenvectors of the covariance matrix Eq. (\ref{Eq:7}).
The so called Principal Component Analysis (PCA) or Karhunen-Loeve transformation \citep{Bishop95} will provide
us with the calculation of the eigenvectors of the covariance matrix of the observation whereby the eigenvectors are
ordered according to their associated eigenvalues (variances). 
We use the PCA method to decompose the entire set of observed Stokes spectra into a new coordinate
system. This procedure will project the most coherent and systematic features in the observed Stokes profiles 
into the first few eigenvectors with the largest eigenvalues while the incoherent features (i.e. noise or blends) will 
be mapped to the less significant eigenvectors (with low eigenvalues). 
After having calculated the set of orthogonal eigenvectors we can use this new basis
to decompose all observed Stokes spectra into the new basis of eigenvectors $\boldsymbol{u_l}$, as 
\begin{equation}
\boldsymbol{x}_k(v) = \sum_l \alpha_{k,l} \boldsymbol{u}_l(v)  \; ,
\label{Eq:8}
\end{equation}
where $\alpha_{k,l} = \boldsymbol{x}_k(v) \boldsymbol{u}_l(v)$ is the scalar product (i.e. the projection or the cross-correlation)
between the observed $k-th$ Stokes profile $\boldsymbol{x}_k(v)$ and the $l-th$ eigenvector $\boldsymbol{u}_l(v)$. 
If we use a limited number of the first few $l$ eigenvectors for this decomposition the reconstruction will be made 
only with those components who carry the most significant information about the systematic features and 
inherent line characteristics.
We therefore reconstruct individual Stokes profiles to an extent that the majority of the individual line characteristics 
are preserved and only a small contribution of uncorrelated effects are present.
To determine the maximum number of meaningful eigenvectors we calculate the covariance matrix of pure noise profiles
with a S/N ratio that corresponds to the noise level of the observations. This allows us to estimate the magnitude of the 
eigenvalues from which the corresponding eigenvectors will mainly contain noise.
Since the PCA multi-line method now allows the reconstruction of \emph{individual} line profiles, all the known 
line parameters of a particular spectral line can be used in a subsequent DI and ZDI inversion.
See \cite{Carroll07,Martinez08} for a more detailed description of the method. 

\subsection{A fast Stokes profile synthesis}
\label{Sect:5.2}
The basic idea of our proposed approach is to emulate the process of 
polarized line formation by using an adaptive regression model, 
that is fast to evaluate, and which provides the required accuracy. 
The adaptive model we seek must provide a sufficient complexity to describe 
the non-linear mapping of Eq. (\ref{Eq:1}), between the most 
prominent atmospheric parameters and the resulting Stokes spectra.
For this purpose, we used a supervised machine learning algorithm, e.g., 
an artificial neural network (ANN) model. A popular type of 
of ANN, also used in this work, is the so called multilayer-perceptron (MLP) which is known as a universal function approximator.
The MLP can be regarded as a class of nonlinear function, which
performs a continuous and multivariate mapping between an input vector $\boldsymbol{x}$ and an output vector $\boldsymbol{y}$.
The network function represents a function composition of elementary
non-linear functions $g(a)$. These elementary functions are arranged in 
layers whereby each of these functions in one layer is connected via an adaptive weight vector $\boldsymbol{w}$ to
all elementary functions in the neighboring layers.
The $l-th$ output (component of the output vector $\boldsymbol{y}$) of a two layer (of weights) MLP for example 
can be written as
\begin{equation} y_{l}(\boldsymbol{x};\boldsymbol{w}) =
g_{k} \left(\sum_{j=0}^{J}w_{kj}^{(2)} g_{j}\left(\sum_{i=0}^{I}w_{ji}^{(1)}x_{i} \right) \right) \; ,
\label{Eq:9}
\end{equation}
where $x_i$ represents the $i$-th component of the input vector $\boldsymbol{x}$ and
$w_{ji}^{(1)}$ the connecting weight from the $i$-th input component to the $j$-th elementary function
$g_j$ in the first unit layer.
The weights $w_{kj}^{(2)}$ then connecting all the functions $g_j$ with the functions $g_k$ in the second unit layer. 
The capital letters (I,J) giving the numbers of elementary functions (units) 
in the respective unit layer.
The elementary functions $g(a)$, which are also called activation functions, are given by the
following type of sigmoid function,
\begin{equation}
g(a) = \frac{1}{1+exp(-a)} \; .
\label{Eq:10}
\end{equation} 
The network function $\boldsymbol{y}(\boldsymbol{x})$  will thus process a given input vector $\boldsymbol{x}$ 
by propagating this vector (via multiplication with the individual weight values and subsequent 
evaluation of the different activation functions) through each layer of the network.
The particular function that will be implemented by the MLP is determined 
by the overall structure of the network and the individual adaptive weight values. 
The process of determining these weight values for the MLP
is called (supervised) training and is formulated as a non-linear optimization process. 
This training is performed on the basis of a representative dataset, 
which includes the input to target (i.e. training output) relations of the underlying problem. 
This process is similar to a non-linear regression for a given
data set, but as the underlying model (i.e., MLP) is much more general, the regression function is 
not restricted to a specific predetermined (or anticipated) model.
In fact, it can be shown that MLPs provide a general framework for approximating 
arbitrary non-linear functions \citep{Bishop95}.
In our case the input-to-target relation is dictated by our
synthesis problem, for the input vectors we have chosen the following atmospheric parameters : 
the temperature and pressure structure of model atmospheres (described by the effective temperature), 
logarithmic gravitation, iron abundance, the local bulk velocity of the plasma, 
microturbulence, macroturbulence, magnetic field strength, magnetic field inclination, 
magnetic field azimuth and the LOS angle between the observer and the local normal.
For the output parameter we have used the full Stokes vector profiles of 
the Zeeman-sensitive iron lines FeI $\lambda$ 6173 \AA\ and FeI $\lambda$ 5497 \AA\,
which have an effective Land\'{e} factor of $g_{\rm{eff}}$ = 2.5 and $g_{\rm{eff}}$ = 2.25 
respectively.
Once the network is successfully trained and has converged in terms of minimizing the error 
between the calculated output vectors and the target vectors of the training database, 
the network weights are frozen and the MLP, which now represents the desired approximation of 
the underlying problem, can be applied to new and unknown data coming from the same parameter 
domain as the training data.
To test the accuracy of the trained MLPs we created a large
number of input vectors with randomly chosen combinations for the atmospheric parameters.
The statistical evaluation shows that the MLPs are able to calculate the corresponding Stokes profiles 
with a high degree of accuracy. The comparison of the MLP calculation with the results from the 
conventional numerical integration of the polarized radiative transfer, performed with the 
DELO method of \cite{Rees89}, demonstrates the impressive results of the MLP synthesis. 
For the Stokes $I$ profile calculation we obtain a rms error as low as 0.11 \%, for Stokes $V$ 0.17 \%; 
and for Stokes $Q$ and $U$, slightly above 1 \% relative to the DELO solution.
The MLP has in fact \emph{learned} to disentangle the different and sometimes
competitive effects of the various atmospheric input parameters to calculate accurate local 
Stokes profiles for both iron lines. 
But it is not only the accuracy of the method which
makes this synthesis method attractive for ZDI inversion, it is first and foremost the speed of the calculation.
A benchmark test with several workstations confirmed that the MLP synthesis is more than a factor 1000
faster then the conventional numerical approach, for more details see \cite{Carroll08a}.    

\section{An artificial neural network approach to ZDI}
\label{Sect:6}
In this section we want to introduce our new statistical inverse approach to ZDI
which is based on artificial neural networks.
Instead of using the ANNs in a forward modeling approach as in the preceding section, we now try to use ANNs to  
provide a direct approximation of the inverse mapping between the 
observed Stokes spectra and the corresponding magnetic surface distribution. 
Again we rely on the non-linear approximation capabilities of the  multilayer-perceptron
as given in Eq. \ref{Eq:9}, to find a continuous approximation of the underlying inverse problem. 
ANNs have been already successfully used in solar Stokes profile inversion problems 
\citep{Carroll01, Carroll08b} which have shown their great potentials in approximating 
the inverse mappings between observed Stokes profiles and the underlying atmospheric parameters. 
One of the key issues in the following ZDI inversion is the reduction of the input (i.e. Stokes spectra)
as well as the output dimension (i.e. surface images) by a PCA decomposition. 
The learning task of the neural network is then to approximate the inverse mapping in the reduced
eigenspace between the eigenprofiles and the corresponding eigenimages of the surface. 
The reduction to the subspace of eigenspectra and images has two main reasons, first, we reduce the 
topology of the MLP structure which allows a more stable and faster convergence in the training process 
and second, the restriction to a subspace of the problem, has a regularizing effect on the solution 
due to the reduced redundancy in the input spectra and the noise reducing effects by describing the solution
as a superposition of eigenimages. 
Because the transformation and training of the network is based upon a training sample, this will have a 
decisive impact on the resulting eigenspectra and eigenimages. A finite training sample
will only provide a good statistical modeling in a limited region of the parameter domain
and an approximation of the inverse function will therefore only provide meaningful results in that restricted domain.
\begin{figure}
 \includegraphics[width=14cm]{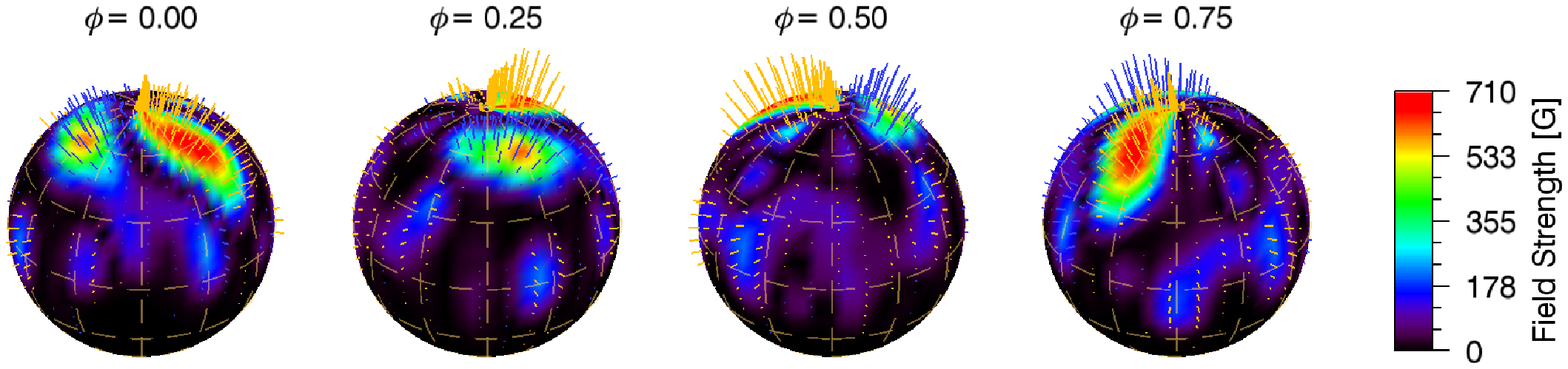}
 \includegraphics[width=14cm]{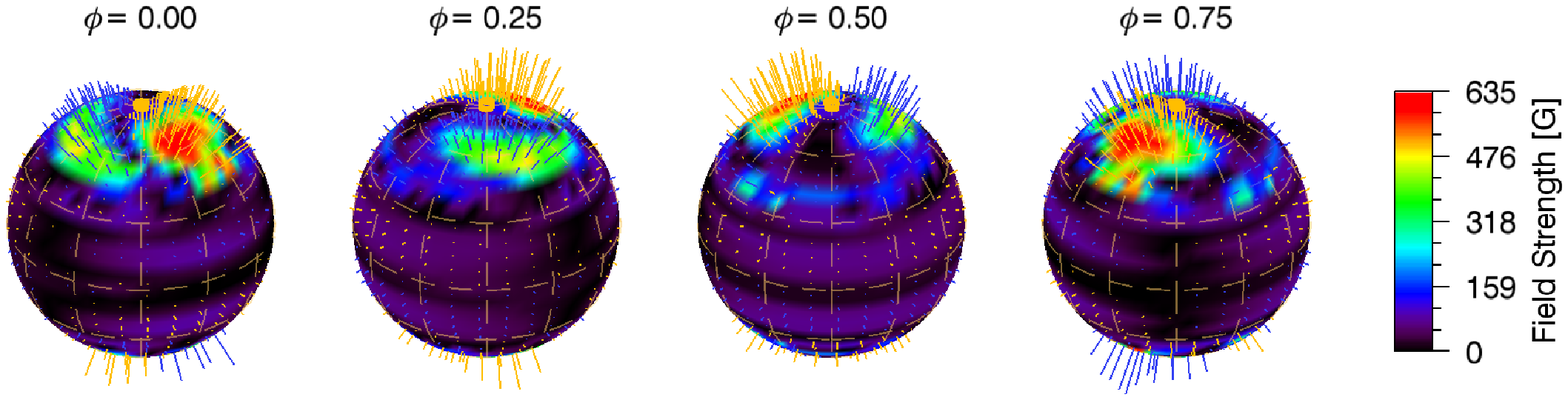}
  \caption{The reconstructed radial magnetic field of II Peg for the year 2007.
  In the top the reconstruction based on the conventional inversion is shown, whereas, the 
  ANN inversion based on the $\alpha^2-\Omega$ dynamo model is shown in the bottom.}  
  \label{Fig:3}
\end{figure}
To apply the ANN inversion on real observed Stokes $V$ spectra \citep{Carroll07}
we trained our MLP on synthesized Stokes profiles which are based on a 
theoretical $\alpha^2-\Omega$ dynamo \citep{Elstner05}. 
The $\alpha^2-\Omega$ dynamo originally developed to study and explain the appearance of 
flip-flop events on active stars provides not only a 
good model representation since our target star II Pegasi is known to be a candidate for an 
flip-flop cycle \citep{Berdyugina07}, it also allows us to incorporate (implicitly)
the physics provided by this model into our inversion approach. From this simulation we 
have synthesized approximately 1000 Stokes $V$ profiles at different phases and 
evolution stages of the dynamo.
Assuming typical stellar and atmospheric
parameters for the subgiant K2 star II Peg \citep{Berdyugina98} 
we have synthesized approximately 1000 disk-integrated Stokes $V$ profiles for different phases and 
different evolution stages of the dynamo simulation.
The training process follows again a large-scale optimization of the network parameters
of the MLP to find the best possible approximation for the inverse mapping between
the decomposed Stokes $V$ profiles and the corresponding decomposed surface images
After the training process the observed II Peg Stokes $V$ spectra applied to the
trained MLP to estimate the magnetic surface structure. 
In Figure \ref{Fig:3} we see a comparison of the surface magnetic field structure
of II Peg retrieved with the conventional ZDI inversion made with our ZDI code \emph{iMap} 
and the new ANN inversion. The surface structure inferred from this first ANN approach 
is in remarkable good agreement with the conventional inversion. The good convergence of 
the training process and the good results of the inversion indicate that the MLP has in fact 
found a well-behaved inverse mapping to that problem and moreover it shows that 
the dynamo model used for this study  provides a reasonable assumption for the surface structure 
observed on II Peg. 
A more detailed description of this novel ANN approach will be given 
in a forthcoming paper \citep{Carroll09}. 

\section{Summary}
\label{Sect:7}
Zeeman-Doppler imaging is an inherent non-linear problem 
even with an appropriate regularization, it is not assured that we
arrive at a global minimum, we are still facing the problem of local minima.
This is particular problematic in combination with an ill-defined or inadequate
forward model. As was shown in Sect. \ref{Sect:4}, even if the observed 
spectra are well reproduced, the solution can be far from the true solution. The
fit to the observed Stokes $V$ profiles does often not provide enough information
to distinguish between competing solutions. 
The additional availability of Stokes $Q$ and $U$ profiles with the next 
generation of spectropolarimeter (PEPSI) at the 8.4 m Large Binocular Telescope (LBT) 
\citep{Strass07,Strass08} will immensely help to further mitigate this problem. 
But for all the current observations where only Stokes $V$ spectra are available
we have to rely on the best possible forward modeling.
In an effort to provide a more systematic radiative transfer modeling we have introduced
a new PCA based multi-line reconstruction technique. This method allows us to recover
Stokes profiles of individual spectra lines and thus facilitates their modeling by means of 
radiative transfer calculations. Moreover, we have introduced a fast Stokes profile synthesis 
which is not only accurate but accelerates the process of calculating the Stokes profiles 
by more than three orders of magnitude.
In a preliminary study we have investigated the capabilities of a new artificial
neural network approach to ZDI. ANNs have a number of favorable features, 
they provide a direct inversion in terms of approximating the inverse mapping between the 
observed Stokes profiles and the magnetic surface distribution.
And moreover, ANNs as a statistical learning tool, allows one to incorporate theory based knowledge 
in the training process which is then utilized in the subsequent inversion. 
Although further investigations are needed, the ANN approach already provides a new and 
promising way of a direct inversion or for a hybrid approach with a conventional ZDI inversion.

\end{document}